\documentclass[preprints,review,accept,moreauthors,pdfTeX]{Definitions/mdpi} 
\firstpage{1} 
\pubvolume{8}
\issuenum{1}
\articlenumber{583}
\pubyear{2022}
\copyrightyear{2022}
\externaleditor{Academic Editors: Luigi Tedesco and Giovanni Marozzi} 
\datereceived{27 September 2022} 
\dateaccepted{27 October 2022} 
\datepublished{4 November 2022} 
\hreflink{https://doi.org/\linebreak 10.3390/universe8110583} 

\usepackage{amsmath,bm}
\usepackage{caption}
\usepackage{subcaption}
\usepackage{tikz} 
\usetikzlibrary{shapes.geometric, arrows}   
\usepackage{amsthm}
\newtheorem*{remark}{Remark}

\usepackage{color}
\definecolor{MyDarkRed}{rgb}{0.7,0,0}

\definecolor{MyBlue}{rgb}{0,40,76}

\newcommand{\dd}{{\hbox{d}}}
\newcommand{\GG}{{\cal{G}}}
\newcommand{\W}{{\cal{W}}}
\newcommand{\A}{{\cal{A}}}
\newcommand{\F}{{\cal{F}}}
\newcommand{\s}{{\cal{S}}}
\newcommand{\CQ}{{\cal{Q}}}
\newcommand{\CD}{{\cal{D}}}
\newcommand{\CW}{{\cal{W}}}
\newcommand{\z}{z}
\newcommand{\y}{y}
\newcommand{\x}{x}
\newcommand{\rv}{\boldsymbol{r}}
\newcommand{\As}{A}

\newcommand{\ltbaverage}[1]{\left\langle #1 \right\rangle_{\rm LTB}}

\Title{On general-relativistic Lagrangian perturbation theory and its non-perturbative generalization}

\TitleCitation{Relativistic Lagrangian perturbations}

\Author{Thomas Buchert $^{1,\ddagger}$*\orcidA{}, Ismael Delgado Gaspar 
$^{2,1,\ddagger}$\orcidB{} and Jan J. Ostrowski 
$^{2}$\orcidC{}}

\AuthorNames{Thomas Buchert, Ismael Delgado Gaspar and Jan J. Ostrowski}

\AuthorCitation{Buchert, T.; Delgado Gaspar, I.; Ostrowski, J.J.}

\address{%
$^{1}$ \quad Univ Lyon, Ens de Lyon, Univ Lyon1, CNRS, Centre de Recherche Astrophysique de Lyon UMR5574, F--69007, Lyon, France ; buchert@ens-lyon.fr\\
$^{2}$ \quad Department of Fundamental Research, National Centre for Nuclear Research, Pasteura 7, 02--093 Warsaw, Poland ; Ismael.DelgadoGaspar@ncbj.gov.pl and Jan.Jakub.Ostrowski@ncbj.gov.pl}

\corres{Correspondence: TB} 

\secondnote{These authors contributed equally to this work.}

\abstract{The Newtonian Lagrangian perturbation theory is a widely used framework to study structure formation in cosmology in the nonlinear regime.
We review a general-relativistic formulation of such a perturbation approach, emphasizing results on already developed extensive formalism including among other aspects:
the non-perturbative modeling of Ricci and Weyl curvatures, gravitational waves and pressure-supported fluids. We discuss subcases of exact solutions related to Szekeres Class II and, as exact average model, Ricci-flat LTB models. This latter forms the basis of a generalization that we then propose in terms of a scheme that goes beyond the relativistic Lagrangian perturbation theory on a global  homogeneous-isotropic background cosmology.
This new approximation does not involve a homogeneous reference background and it contains Szekeres class I (and thus general LTB models) as exact subcases. Most importantly, this new approximation allows for the interaction of structure with an evolving `background cosmology', conceived as a spatial average model, and thus includes cosmological backreaction.
}

\begin{document}



\section{Introduction}
Relativistic cosmological perturbation theory gave rise to many exciting results regarding the early Universe, CMB (Cosmic Microwave Background) physics, and the late Universe for the formation of large-scale structure. General relativity, unlike other physical theories describing the weak and strong nuclear forces and electrodynamical interactions, does not rely on the notion of \textit{background}, understood as a pre-fixed absolute spacetime. This fact together with the nonlinearity of general relativity introduces conceptual difficulties in the formulation of a cosmological perturbation theory. Difficulties that can be at least partially circumvented in the astrophysical context by a suitable choice of boundary conditions, e.g. asymptotic flatness, but not when we aim at describing the Universe as a whole.

Whether we consider the inflationary epoch, scalar fields and semi-classical quantum gravity (see e.g. \cite{Brustein95}, \cite{Gasperini1997}) or the matter-dominated, fully classical Universe, we can usually distinguish two main approaches to relativistic cosmological perturbations, based on which quantity is being perturbed. In both cases we have to put a reference background by hand and the commonly adopted choice is a member of the family of FLRW (Friedmann-Lema\^\i tre-Robertson-Walker) spacetimes. The original work by Lifshitz \cite{lifshitz46} 
(see also \cite{lifshitzkhalatnikov}, \cite{hawking}, \cite{tomita} and follow-up papers of this author) introduced perturbations of the FLRW metric coupled to the perturbations of the stress-energy tensor via the linearized Einstein equations. This work employs intrinsic comoving coordinates. The issue of coordinate choices later highlighted the problem of gauge fixing, i.e. choosing the way of mapping the physical spacetime to the assumed background, which can be thought of as a mapping between two distinct manifolds in the framework of \textit{gauge-invariant perturbation theory} \cite{mukhanovetal,kodamasasaki,MalikWands2009,Durrer}. It is worth noticing that the gauge fixing allows to pullback the physical quantities from the physical spacetime, resulting in e.g. curvature perturbations residing as fields on the background manifold rather than as intrinsic geometrical features of the physical manifold. Such a set-up is in the spirit of the Eulerian picture in hydrodynamics where we examine the fluid in specific, fixed locations as opposed to the Lagrangian picture where we follow the fluid flow intrinsically \cite{Ehlers61}. In this context, the background spacetime defines an arbitrary, absolute reference frame. In the seminal paper \cite{Bardeen1980}, gauge-invariant variables where defined. However, although the basic physical perturbation variable is the density contrast, none of the Bardeen's variables actually represented this quantity. This issue was partially resolved by John Stewart \cite{stewart}. A different (although fully compatible) approach was presented in \cite{EllisBruni1989}, where $1+3$ threading of spacetime was utilized. It can be considered as a step towards the Lagrangian approach to relativistic cosmological perturbation theory despite the fact that the mapping between two manifolds was still in place. However, in the above-mentioned work, the authors define the variables related to the congruence of the world-lines of the fluid (spatial projections of density gradient and extrinsic curvature) and the mapping exists between the fluid trajectories not between fixed points in the two manifolds.

While gauge-invariant variables have to be constructed afresh at any order of the perturbations, a genuine general-relativistic treatment does not suffer from this issue when we consider trace-free tensors that vanish on the chosen reference background \cite{stewart}. We shall be guided by covariant formulations that do not rely on the choice of coordinates, but on the choice of a foliation of spacetime \cite{foliations,buchert:generalfluid}. Gauge choices in the two-manifold approaches can be mapped to foliation choices in a covariant setting, see \cite{vitentietal} and references therein (e.g. the series of papers by Mario Novello and collaborators). The Lagrangian coordinates, being intrinsic to the fluid congruence, allow us to keep covariance on spatial leaves of a $3+1$ foliation, while the existence of hypersurfaces allows us to define a background evolution for scalar variables that are conceived as spatial averages of geometric and matter variables \cite{generalbackground}.

Before moving to a more detailed description of Lagrangian cosmological perturbation theory (Newtonian and relativistic), we wish to recall the fundamental problem signalized at the beginning of this introduction: general relativity does neither rely on the notion of background, nor does it give us \textit{a priori} any prescription on how to construct one \cite{combi}. One of the aspects of this ever persistent conundrum is the fact that the FLRW background metric and the spatially homogeneous density obtained by any averaging or smoothing procedure involving limits (e.g. integrating or taking the limit of some parameter related to scale or inhomogeneities) are, by construction, not related to each other through the Einstein equations. The first part of the problem becomes apparent when we try to define the average of a tensor which without any additional assumptions is an ill-defined concept. Even if we had defined such operation it would not commute with taking the inverse of the metric and calculating the curvature (due to the quadratic terms), and this would necessarily lead to an additional term in the Einstein equations, the backreaction term. We note that already volume-averaging and time-evolution are non-commuting operations, and their difference consists of positive-definite backreaction terms even before invoking dynamics or a theory of gravity  \cite{buchert:average,EllisBuchert2005}. This also holds for operations such as Ricci-flow smoothing and scaling in a 3-manifold \cite{buchertcarfora1} and for other coarse-graining techniques such as renormalization \cite{carfora:renormalization}.
The perhaps simplest and to date widely used approach to the backreaction problem consists in scalar averaging of  Einstein's equations, within frameworks of $3+1$ foliations \cite{buchert:dust, buchert:fluid, buchert:generalfluid}, $2+1+1$ foliations on the lightcone \cite{buchert:lightcone}, and their manifestly 4-covariant formulations \cite{Gasperini2, Gasperini3}, see also \cite{foliations, covariance}.  Although essential to the relativistic perturbation theory, the topic of backreaction lies beyond the scope of this review except for the part where we describe how a concrete model for inhomogeneities such as the Lagrangian perturbation theory can serve as a closure condition to the otherwise under-determined system of averaged equations. 
 It was also not our ambition to give an exhaustive account of all of the many developments in cosmological perturbation theory in general. We rather focus on developments in relativistic Lagrangian perturbation theory, starting however with a brief introduction into Newtonian Lagrangian perturbation theory.

\subsection{A brief history of Lagrangian perturbation theory for Newtonian flows}

The context that led Yakob Zel'dovich to his proposal
of extrapolating the Eulerian linear theory of gravitational instability further into the
nonlinear regime may be traced back to a book project he conducted together with Anatoly Myshkis, in which inertial motion and the formation of structure was discussed in affectionate spadework \cite{zeldovichmyshkis:book}. 
Also, Zel'dovich might have been aware of related work in plasma physics such as Dawson's discussion of multi-streaming in a cold plasma \cite{Dawson1959}.
In view of this, he must have seen the
superiority of a Lagrangian description of fluid motion over the established Eulerian perturbation theory for the purpose of understanding and describing structure formation
in the Universe. Upon linearizing the exact solution for Galileian continua, 
it is evident that a time-dependent rescaling of the dependent and independent variables 
suffices to match both the Eulerian linear perturbation solution on a Friedmannian background cosmology (see, e.g. \cite{peebles1:book}) and the powerful nonlinear solution for inertial systems. His early papers \cite{zeldovich70a,zeldovich70b,zeldovich78}, \cite{zeldovichshandarin},
already explore in detail the consequences of this idea, which -- contrary to popular opinions at the time 
(nonlinear evolution was understood, both analytically and numerically, through 
spherically symmetric systems) -- predicts a highly anisotropic process of structure formation 
into high-density objects. He called the first collapsing objects \textit{bliny}, the Russian word for
small \textit{pancakes}.
He had in mind the local picture which, later, was `americanized' to extended Megaparsec-\textit{pancakes}.

Soon thereafter, confirmations of this idea in the context of
the Euler-Newton system for self-gravitating dust (i.e., pressure-less matter) were advanced within Zel'dovich's group. The paper \cite{doroshkevichetal73} (among others, see e.g. the considerations on the density probability distribution in \cite{doroshkevich70}, \cite{grinsteinwise8789}, \cite{bartelmannschneider92}, investigations on self-similar solutions \cite{fillmoregoldreich} and trajectories \cite{bartelmann}.) 
stands out as a convincing analytical justification. The heart of this
work is a test of self-consistency of \textit{Zel'dovich's approximation}, henceforth denoted ZA, conducted through comparison of the density from (i) the field equations and (ii) from the exact density integral (this test is explicitly discussed in \cite{buchert89}). As a side-result this test also shows that Zel'dovich's approximation 
provides a special exact solution for plane-symmetric motions on a Friedmannian background. In parallel, Sergei Shandarin
performed a numerical simulation, 
at low resolution at the time (see references in \cite{shandarinzeldovich89}), but confirming the above picture, and reinforcing what was 
earlier found by Lin, Mestel and Shu \cite{linmestelshu} 
(see also \cite{goodmanbinney}, \cite{ZAaccuracy}). 

Further numerical work within Zel'dovich's group \cite{doroshkevichetal80}, 
\cite{klypinshandarin}, and especially
Sergei Shandarin's move to Lawrence, Kansas, initiated detailed
numerical investigations of the \textit{pancake picture} together with Adrian Melott, who developed
his expertise in the early 80's 
\cite{melottshandarin89}. 
The ``Eastern'' \textit{pancake picture}
was by then popular for massive neutrino-dominated (\textit{top-down structure formation}) cosmogonies, 
in which structure on large scales forms first and later fragments into smaller units. It was 
an equal competitor to the ``Western''
(\textit{bottom--up structure formation})
hierarchical cosmogony,
in which large-scale structure
forms out of small-scale inhomogeneities in the course of a merger process. The nowadays still accepted $\Lambda$CDM standard model of cosmology performs in between these two scenarii. 

Returning to the history of Zel'dovich's analytical model, the collaboration with Vla\-dimir 
Arnol'd was probably the final highlight for inhomogeneous dust matter models \cite{arnoldshandarinzeldovich},
in which the \textit{Lagrange-singularity theory} \cite{arnold:book} was 
employed to furnish a classification scheme of the building blocks of large-scale structure
in the Universe. It came right after two further analytical pieces that were missing from the
picture: the form of Zel'dovich's approximation for the modeling of structure formation on Friedmannian background cosmologies with 
curvature \cite{shandarin80}, and the \textit{derivation} 
of exact plane-symmetric solutions for self-gravitating dust flows \cite{zentsovachernin}, 
confirming what was suggested by \cite{doroshkevichetal73}\footnote{For further discussion of and references to the singularity problem see Ref. \cite{rza1}, Section VB.}. The effect of velocity dispersion was discussed by Shukurov \cite{shukurov}. Thereafter, an extension of Zel'dovich's approximation to phenomenologically access the multi-stream regime was proposed in \cite{gurbatov89} through application of the known solution to the 3D Burgers equation: a Laplacian forcing was added to the scaled inertial system. A derivation of this \textit{adhesion approximation} from kinetic theory was provided in \cite{adhesion}; see the review \cite{adhesive} that also investigates deviations from mean field gravity.\footnote{In this context it is important to remark that the best numerical methods are simulations in phase space that provide a controlled and detailed access of the formation of structure via a projection into space \cite{bertschinger:phasespace}. Since the phase space is six-dimensional this, however, provides limits of resolution. For recent simulations see 
\cite{Abel1,Abel2} and the review \cite{Hahn:review}. For a discussion of eigenvalues of the deformation tensor, see \cite{sharvari3}.}
A further extension was suggested in \cite{extending} by including a noise term to lead to KPZ-type equations.
For a summary of the different approximations in a cosmological context, including non-perturbative approaches and exact integrals of the field equations, see \cite{nonperturbative}.

Most of these early achievements and a wealth of application work of Zel'dovich's approximation were done before
a \textit{Lagrangian theory for self-gravitating flows} and, based on this, a \textit{Lagrangian
perturbation approach} was even formulated;
the basic mathematical framework 
(apart from the self-consistency test in  \cite{doroshkevichetal73}
and the plane-symmetric investigation in \cite{zentsovachernin}) was still that of inertial motion. Final confirmation and derivation of Zel'dovich's approximation were based
on the transformation of the full Euler-Newton system to its Lagrangian form 
\cite{buchertgoetz}, \cite{buchert89}, \cite{bildhaueretal}, and, finally, the systematic framework of 
a Lagrangian perturbation theory, in which Zel'dovich's approximation forms a subclass
of the first-order solutions \cite{buchert89}, \cite{buchert92}; closed form solutions at second \cite{buchertehlers}, third \cite{buchert94} and fourth order \cite{rampfbuchert} can be reconstructed using the n-th order perturbation and solution schemes investigated in \cite{ehlersbuchert} (see also more recent work on recursion relations \cite{matsubara2,rampf:recursion} and related investigations of time-analyticity or smoothness of trajectories \cite{Zheligovsky2014}, \cite{Rampfetal2015}). The early 90's can be considered as the birth of Lagrangian perturbation theory, when other groups started to contribute
\cite{moutarde}, \cite{bouchetetal92,bouchetetal95} (see the early review papers \cite{shandarinzeldovich89}, \cite{sahnicoles}, 
\cite{ehlersbuchert},  a brief summary \cite{buchert93L}, the tutorial \cite{buchert_varenna}, the more general review \cite{bertschinger:review}, 
and the more recent review \cite{bernardeau:review}).
Most of these investigations employ the matter model dust; velocity dispersion can be treated as an anisotropic pressure
that adds a Laplacian forcing as in the adhesion approximation \cite{gurbatov89,adhesion}. In a Lagrangian perturbation approach this Laplacian forcing must be transformed to Lagrangian coordinates, which introduces substantial nonlinearity to the problem \cite{adlerbuchert}.

\begin{figure}[ht]
\begin{center}
\includegraphics[width=0.7\textwidth]{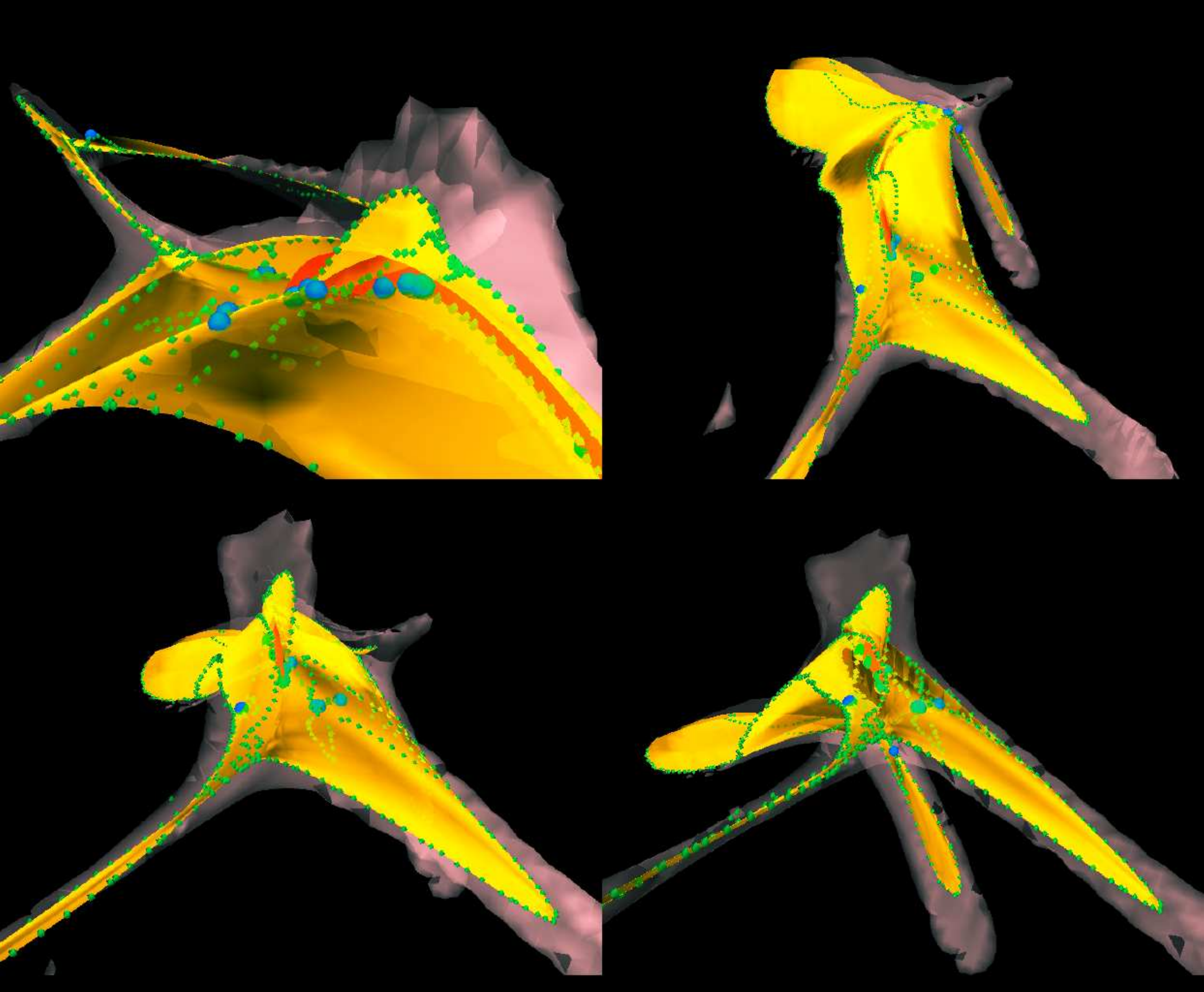}
\end{center}
\caption{\label{fig:pancakes} 
An illustration of the complexity of caustic formation that already arises in a three-dimensional
inertial system. Shown are four views of the same evolution stage. Yellow surfaces correspond to caustic surfaces from the degeneracy of the first eigenvalue of the deformation tensor; red surfaces correspond to secondary degeneracies of the second eigenvalue; third degeneracies are barely visible and may occur later. A hierarchy of further nested caustics forms by including gravity. Green dots correspond to \textit{cusp 
singularities}, while the blue dots correspond to 
\textit{higher-order singularities}. Pink-grey shaded areas correspond to density contours. Figure by Arno Weiss \cite{bsw}.}  
\end{figure}

We remark that current simulations of structure formation are initialized by the Zel'\-dovich approximation or its higher-order extensions. What is not well-known is that this approximation and higher-order extensions also perform well in comparison with numerical simulations until the present epoch of structure formation, if the initial spectrum is appropriately truncated to avoid overproduction of structure after shell-crossing \cite{tza,melottpellmanshandarin},\cite{buchertmelottweiss,melottbuchertweiss,weissetal96}; for the adhesion 
 approximation \cite{gurbatov89}, see \cite{weinberggunn}; for recent direct comparisons between higher-order Lagrangian perturbation solutions and numerical simulations, see \cite{Michaux2021,Schmidt2021}. Shell-crossing itself was also in the focus of interest \cite{buchertehlers,buchert_performance,yanoetal}, a sustained interest up to the present day, see e.g. the recent papers \cite{rampffrisch,colombi:shellcrossing1,colombi:shellcrossing2,rampf:shellcrossing1,rampf:shellcrossing2,rampf:vlasov,tyger}. Despite shell-crossing, Lagrangian schemes are capable of quickly simulating a large number of realizations of Hubble volumes, a task that is still executed with expensive numerical simulations. This remark is especially relevant, since current resolution power of numerical simulations still falls short to access small-scale structure within Hubble volumes, a scale that roughly corresponds to scales where also the analytical schemes fail. 
 The simulation of mock catalogs is an exception, see e.g. \cite{monaco1,monacorev,monaco2}, \cite{kitaura1,kitaura2} and \cite{ZAnumerics} and references therein. The corresponding general-relativistic schemes that we shall introduce would in addition allow to map the curvature distribution in mock catalogs, and from the metric we will be able to implement concrete ray-tracing methods along the lightcone for a precise estimation of lensing effects and for realistic estimates of distances.
 
 Analytical schemes allow for extremely high-resolution studies. After the first study in two spatial dimensions 
 \cite{buchert89}, further insights were obtained for three-dimensional special and generic initial data \cite{buchertbartelmann,buchert_performance}
 including hints for a mechanism of gravitational fragmentation at higher orders in Lagrangian perturbation theory.\footnote{
In order to achieve even higher resolution, it is possible to interpolate initial data, since they are smooth and the analytical scheme maps any (calculated or interpolated) point to the nonlinear stage.}
 Detailed studies of 3D pancakes have been conducted, but did not end up in publications at the time, an example is shown in Figure \ref{fig:pancakes}.\footnote{Recently, using newer numerical methods, \cite{hiddingetal} provided a detailed generalization of the 2D pancakes of \cite{arnoldshandarinzeldovich}.}  

During the following decades, many papers referring to Lagrangian approaches have been published, also guided by 
improvements of numerical technologies.
These works initiated a revival of Lagrangian perturbation theory at the beginning of this millenium: the inverse Lagrangian perturbation solutions have been employed for reconstruction techniques, see \cite{recon} and references therein to earlier work, \cite{courtois:reconstruction} and follow-up papers of these authors, as well as related techniques \cite{frisch:reconstruction}. Work is pursued, e.g. for convergence issues \cite{sharvari1,rampf:shellcrossing1}, re-expansions \cite{sharvari2}, renormalization, re-summation and recursion techniques \cite{crocce}, \cite{matsubara1}, \cite{matsubara2,rampf:recursion}, respectively, redshift space distorsions \cite{hivon}, \cite{white} and references therein,
and statistical biasing relations between the galaxy and matter distributions \cite{biasing} (a certainly incomplete list).

\subsection{Relativistic generalization of Lagrangian perturbation theory}

For the introduction of a dictionary of Newtonian variables and equations to general relativity, we first recall the governing equations in Newtonian theory and restrict these considerations to a self-gravitating dust continuum (i.e. a pressure-less fluid). 

\subsubsection{Lagrange-Newton system}

The description of fluid motion in Newtonian cosmology and in particular the problem of perturbations
of the Hubble flow have been largely studied 
in an \textit{Eulerian} coordinate system that is comoving with the Hubble flow, ${\bf x} = a(t){\bf q}$, with $a(t)$ a solution in the family of FLRW cosmologies; ${\bf x}$ are 
inertial non-rotating coordinates in Euclidean space. 
On the contrary, the Lagrangian coordinate system, denoted by $\bf X$, 
is a curvilinear, possibly rotating frame in Euclidean space which is defined such as to move 
with the fluid elements. The Lagrangian description enjoys a number of advantages over
the Eulerian one, some of which we shall highlight. This remains true within a perturbation approach. For an early nice introduction into the Lagrangian framework, see \cite{serrin}, for a systematic introduction for self-gravitating Newtonian flows, see \cite{ehlersbuchert}. For many complementary investigations of relativistic Lagrangian perturbations see, e.g. \cite{ornella1,ornella2}, \cite{terranova}, \cite{kasai93,russ1,russ2}, including gradient expansions \cite{gradientexpansion1,gradientexpansion2} and follow-up papers by these authors, as well as the recent papers on gradient expansions \cite{Rigopoulos2012} and \cite{Rampf2013}. 

Summarizing the Lagrange-Newton system briefly, we note that the one-parameter family of diffeomorphisms, ${\bf X} \mapsto {\bf x} = {\bf f}({\bf X},t)$ allows to reduce Eulerian fields to definitions or provide exact integrals for them, e.g. the velocity field ${\bf v}$, the acceleration field (equivalent to the Newtonian field strength ${\bf g}$ for a dust matter model), the density $\varrho$, and the vorticity $\boldsymbol{\omega} = \nabla \times {\bf v}$, can be derived from the knowledge of the family of trajectories ${\bf f}$:
\begin{equation}
\mathbf{x}:= \mathbf{f}(\mathbf{X},t) \;;\; \mathbf{v}:= \dot{\mathbf{f}}(\mathbf{X},t) \; ;\; \mathbf{a}:= \ddot{\mathbf{f}}(\mathbf{X},t) \;;\;
\varrho(\mathbf{X},t)=\frac{\mathring{\varrho}}{J(\mathbf{X},t)} \;;\; \boldsymbol{\omega} = \frac{\mathring{\boldsymbol{\omega}}\cdot \boldsymbol{\nabla}_{\bf 0} \mathbf{f}}{J(\mathbf{X},t)}\;\,; \;etc. \, ,
\label{identi}
\end{equation}
where $\mathring{\varrho}$ and $\mathring{\boldsymbol{\omega}}$ stand for the initial values of the density and vorticity fields,
$\boldsymbol{\nabla}_{\bf 0}$ for the nabla operator with respect to the Lagrangian coordinates, and $J$ is the determinant of the Jacobian matrix $(f^{a}_{\; \vert j})$,
\begin{equation}
J := \det (f^a_{\; \vert j}) (\mathbf{X},t) = \frac16 \epsilon_{ijk}  \epsilon^{klm} f^i_{\ \vert k}f^j_{ \ \vert l} f^k_{\ \vert m}\:\:,
\end{equation}
where we have denoted by a vertical slash ${}_{\vert i}$ the spatial derivative with respect to $X_i$, by an overdot the partial time derivative, $ \epsilon_{ijk} $ being the Levi-Civit\`a symbol.\footnote{We have distinguished counter indices $a,b,c, \cdots$, here counting vector components, from coordinate indices $i,j,k, \cdots$. We shall later see that this is convenient when a coordinate basis does not exist; here the vector components are also coordinate indices in the Eulerian basis.} 
The remaining field equations of the Euler-Newton system,
$\boldsymbol{\nabla}\times  \mathbf{g} = \mathbf{0}\ ; \boldsymbol{\nabla}\cdot \mathbf{g} =  \Lambda - 4 \pi G \varrho$, become evolution equations in the Lagrangian picture, comprising the closed \textit{Lagrange-Newton system}:
\begin{eqnarray}
\label{LNS1}\delta_{ab} \ddot{f}^{a}_{\;\,\vert [i}\,  {f}^{b}_{\;\,\vert j]} = 0\quad;\quad \frac{1}{2} {\epsilon}_{abc} {\epsilon}^{ikl}\ddot{f}^{a}_{\;\,\vert i}  \, f^b_{\;\,\vert k} \, f^c_{\;\,\vert l} = \Lambda J - 4 \pi G \mathring{\varrho} \;.
\end{eqnarray}
We notice that this is a system of equations for the gradient ${\bf d} f^a$, i.e. involving matrices for the derivatives of $\bf f$ that can be easily generalized to tensors as we shall see below.

A Lagrangian perturbation theory results from the ansatz that represents ${\bf d}f^a$ in terms of a homogeneous-isotropic deformation and expands the Lagrange-Newton system to a given order in the deviation gradient field ${\bf d}{\bf P}$: ${\bf d}{\bf f} = a(t)({\bf X} + {\bf d}{\bf P})$.   

\subsubsection{Lagrange-Einstein system}

Moving to general relativity, we have to give up the global vector space that is needed to construct vector perturbations and their gradients. Instead, this global vector space reduces to the tangent space at a point on the Riemannian manifold.
Thinking of Riemannian space sections, we introduce a 
{\it spatial} metric form $\bf g$ with coefficients $g_{ij}$ in an exact (co-tangential) basis ${\bf d}X^i \otimes {\bf d}X^j$. We can write any such metric as a quadratic form of deformation one-forms or co-frame fields, ${\bf g} = \delta_{ab} \,\boldsymbol{\eta}^a \otimes \boldsymbol{\eta}^b$, i.e. in terms of coefficients, $g_{ij} = \delta_{ab} \eta^a_{\;\,i}\eta^b_{\;\,j}$. Such a metric form is \textit{Ricci-flat}, i.e. its Ricci tensor vanishes everywhere, if there exist functions $f^a$, such that the one-forms can be written as exact forms,
\begin{equation}
\label{dictionary}
\boldsymbol{\eta}^a \mapsto \boldsymbol{\mathrm d} f^a \ .
\end{equation}
We call \eqref{dictionary} the \textit{Euclidean restriction}. 
It can be profitably used to either extrapolate Newtonian fields written in Lagrangian form into general-relativistic expressions through inversion of \eqref{dictionary}, or it transforms any general-relativistic (here spatial) expression into its Newtonian counterpart. This 
provides a straightforward dictionary that, e.g. in a flow-orthogonal foliation of spacetime of a dust matter model, does not need to send the speed of light to infinity as in standard Newtonian limits \cite{bm}. 
We note that the Eulerian form of the Newtonian 
equations, written in an inertial coordinate system, depends on these coordinates; to cure this coordinate-dependence, we can transform to rotating coordinates and collect the non-inertial forces in connection coefficients (the Newton-Cartan equations, e.g. \cite{QuentinNC} and references therein). A connection also arises through the restriction of the spatial relativistic Levi-Civit\`a connection using \eqref{dictionary}, leading to an integrable Newtonian connection in a Lagrangian frame \cite{rza1}. 
For any metric, if we can find a coordinate transformation $x^i = f^{a\equiv i} (X^j , t)$ that transforms the Euclidean metric coefficients, $\delta_{ij} {\mathrm d}x^i {\mathrm d}x^j = \delta_{ab} f^a_{\;\,|i} f^b_{\;\,|j} {\mathrm d}X^i {\mathrm d}X^j$, into the metric coefficients $g_{ij}= \delta_{ab} f^a_{\;\,|i} f^b_{\;\,|j}$, then these latter are just a rewriting of the flat spatial metric. Given this remark, any perturbation theory that features metric forms of the integrable form, does not describe relativistic inhomogeneities; metric coefficients of the above form describe Newtonian (Lagrangian) perturbations on a flat background space. An intrinsic general-relativistic perturbation theory deforms the background geometry; in other words, the perturbations live in a perturbed space, not on a reference background. This remark also shows that perturbation terms can be identified through their integrability in the spirit of \eqref{dictionary} to be coordinate artifacts. We shall again encounter these thoughts when we discuss gravitational wave solutions. 

The foundational paper I, \cite{rza1}, and the subsequent series of papers, transforms the following Einstein equations for the case of irrotational dust in a $3+1$ representation, i.e. the $3$ irrotationality conditions on the expansion tensor $\Theta_{ij}$ with its trace denoted by $\theta$, the $6$ evolution equations for the $6$ components of the symmetric spatial metric $g_{ij}$ with the $4$ energy and momentum constraint equations:
\begin{subequations}
\label{einstein1}
\begin{align}
\Theta_{[ij]} = 0 \; ; \\
{\dot g}_{ij} = 2\,g_{ik}\Theta^k_{\;j}\quad ; \quad 
{\dot \Theta}^i_{\;j} + \theta \Theta^i_{\;j} = -{\cal R}^i_{\;j} + (\,4\pi G \varrho + \Lambda\,)\,\delta^i_{\;j}\; ;\\
\theta^2 - \Theta^i_{\;j}\,\Theta^j_{\;i} = - {\cal R} + (16\pi G \varrho + 2\Lambda)\; ;\\
 \Theta^i_{\;j|| i} - \theta_{|j} = 0\; ,
\end{align}
\end{subequations}
where ${\mathcal R}_{ij}$ denote the coefficients of the spatial Ricci tensor, with ${\mathcal R}$ its trace; 
an overdot denotes a partial derivative with respect to the coordinate time (being equivalent to the covariant time-derivative in the present setting); a double vertical slash denotes the covariant spatial derivative with respect to the $3$-metric, the connection is assumed to be Levi-Civit\`a; the dust density is given by $\varrho = \mathring{\varrho} J^{-1}$, $J \ge 0$; $J$ defines the coefficient function of the $3-$volume form, $J \equiv {\sqrt{g}} / {\sqrt{G}}$, with
${\sqrt{g}} \ {\bf d}^3 X$ the $3$-volume form on the exact basis, $g: = \det (g_{ij}(\mathbf{X},t))$ and $G: = \det (G_{ij})= \det (g_{ij}(\mathbf{X},t_{\rm i}))$.

The system \eqref{einstein1} is transformed into  
a set of $13$ equations for the three spatial co-frame fields, i.e. $9$ evolution equations for the $9$ components of the spatial co-frames and $4$ equations stemming from the $4$ constraint equations in the classical metric representation \eqref{einstein1}.
We call the resulting system the \textit{Lagrange-Einstein system}:\footnote{Starting with Paper II in the series, we employ the metric form coefficients $g_{ij} = G_{ab} \eta^a_{\;\,i}\eta^b_{\;\,j}$, using Gram's matrix $G_{ab}$ instead of $\delta_{ab}$ for orthonormal co-frames. It has the advantage that the initial deformation can be set to be undeformed and therefore provides a closer correspondence to Newtonian solutions. The initial metric is then not encoded in initial deformations but in Gram's matrix (for details see \cite{rza2,rza3}; for the Euclidean restriction (integrable limit) of Gram's matrix see [appendix A]\cite{rza3}), and for the integrable limit of the gravito-magnetic part see [Sect. III.A.2]\cite{rza1}. Note also that Equations \eqref{form_symcoeff} imply $G_{ab} \,\ddot{\eta}^a_{[i} \eta^b_{\ j]} = 0 \ $.}
\begin{subequations}
\label{LES}
\begin{align}
\label{form_symcoeff}&G_{ab} \,\dot{\eta}^a_{[i} \eta^b_{\ j]} = 0 \;; \\
\label{form_eomcoeff}&\frac{1}{2 J} \epsilon_{abc} \epsilon^{ikl} \left( \dot{\eta}^a_{\ j} \eta^b_{\ k} \eta^c_{\ l} \right) \dot{} = -{\mathcal R}^i_{\ j} + \left( 4 \pi G \varrho + \Lambda \right) \delta^i_{\ j}\;; \\
\label{form_hamiltoncoeff}&\frac{1}{J}\epsilon_{abc} \epsilon^{mjk} \dot{\eta}^a_{\ m} \dot{\eta}^b_{\ j} \eta^c_{\ k} = - {\mathcal R}+ \left( 16\pi G \varrho + 2\Lambda  \right) \;; \\
\label{form_momcoeff}&\left(\tfrac{1}{J}\epsilon_{abc} \epsilon^{ikl} \dot{\eta}^a_{\ j} \eta^b_{\ k} \eta^c_{\ l} \right)_{||i} = \left(\tfrac{1}{J}\epsilon_{abc} \epsilon^{ikl} \dot{\eta}^a_{\ i} \eta^b_{\ k} \eta^c_{\ l} \right)_{|j} \; .
\end{align}
\end{subequations}
In this system we left the Ricci tensor and its trace as well as the covariant derivative implicit, since writing them out in terms of the co-frames becomes non-transparent for our discussion. 

We can recover the previously presented Lagrange-Newton system through elimination of the Ricci curvature. We keep the first $3$ equations \eqref{form_symcoeff} that also hold for the second time-derivative of the co-frame fields, and Raychaudhuri's equation (derived from the trace of the equation of motion \eqref{form_eomcoeff} combined with the energy constraint \eqref{form_hamiltoncoeff}):
\begin{equation}
\label{LESgravitoelectric}
G_{ab} \,\ddot{\eta}^a_{[i} \eta^b_{\ j]} = 0 \quad;\quad
\frac{1}{2J} \epsilon_{abc}\epsilon^{ik\ell} \ddot{\eta}^a_{\ i} \eta^b_{\ k} \eta^c_{\ \ell} = \Lambda - 4 \pi G {\varrho} \;.
\end{equation}

The dictionary \eqref{dictionary} works and reproduces what we call the \textit{gravito-electric part} of the Lagrange-Einstein system consisting of $4$ equations.
Its Euclidean restriction is a closed system (the Lagrange-Newton system \eqref{LNS1}) for the three 
vector components $f^a$ of the Lagrangian deformation.
The $9$ components of co-frame fields are solutions of this system complemented by the \textit{gravito-magnetic part} of the Lagrange-Einstein system. Perturbation solutions are tensorial and feature a richer structure within the full system. We can trace this back to the fact that after elimination of the trace of the Ricci tensor we are left with 
Equation \eqref{form_eomcoeff} for the trace-free part of the Ricci tensor. An evolution equation for trace-free parts is not needed to find Newtonian solutions, e.g. the Newtonian tidal tensor can be represented in terms of solutions of \eqref{LNS1}.

\subsection{Summary of results on relativistic Lagrangian perturbations}

We now discuss the main contents of the series of papers following Paper I \cite{rza1}. 

\noindent
We only briefly mention Paper III \cite{rza3} in the series that gives the n-th order perturbation and solution schemes, where emphasis is on the gravito-electric part inspired by the corresponding perturbation and solution schemes in the Lagrange-Newton framework \cite{ehlersbuchert}.
We discuss more in detail the results of Paper II \cite{rza2} in the series that invokes spatial averaging of the first-order solutions and provides an explicit form of the \textit{backreaction functional}. We give a brief account of the basic notions of spatial averaging and the main results of Paper II in what follows. 

Spatially averaging Einstein's equations, we can write the result in terms of Friedmann-like equations for a volume scale factor $a_\CD : \propto V_\CD^{1/3}$, suitably normalized, that appear to be sourced by an 
\textit{effective} perfect fluid energy-momentum tensor \cite{buchert:dust,buchert:fluid}, reviews are \cite{buchert:review,buchert:focus,buchertrasanen}:
\begin{subequations}
\begin{eqnarray}
3\frac{{\ddot{a}}_{\CD}}{a_{\CD}}  =  - 4\pi G(\varrho_{\rm eff}^{\CD}+3{p}_{\rm eff}^{\CD})+\Lambda \;;\;\nonumber\\
3H_{\CD}^{2}- \frac{3 k_{\CD}}{a_{\CD}^2}= 8\pi G\varrho_{\rm eff}^{\CD}+\Lambda \;;\; \nonumber\\
{\dot{\varrho}}_{\rm eff}^{\CD}+3H_{\CD}(\varrho_{\rm eff}^{\CD}+{p}_{\rm eff}^{\CD})=0\,,
\label{eq:effectivefriedmann}
\end{eqnarray}
where the effective densities are defined as 
\begin{eqnarray}
&&\varrho_{{\rm eff}}^{{\CD}} := \langle{\varrho}\rangle_\CD + \varrho_{\Phi}\;\;\;;\;\;\varrho_{\Phi}  :=  -\frac{1}{16\pi G}{\CQ}_{{\CD}}-\frac{1}{16\pi G}\CW_\CD \ ;\nonumber
\label{eq:equationofstate}\\
&&{p}_{{\rm eff}}^{{\CD}} := p_{\Phi}\;\;\;;\;\;\, p_{\Phi} :=  -\frac{1}{16\pi G}{\CQ}_{{\CD}}+\frac{1}{48\pi G}\CW_\CD \ .
\end{eqnarray}
\end{subequations}
In this form the effective equations suggest themselves to interpret the extra fluctuating sources in terms of a scalar field (the morphon field \cite{morphon}), which refers to the inhomogeneities in the geometric functionals $\CW_\CD$ and $\CQ_\CD$.\footnote{This rewriting of the backreaction terms in the form of a minimally coupled effective scalar field can be profitably used to establish quintessence models without dark energy, or inflationary models without fundamental scalar field \cite{morphoninflation}, or models mimicking dark matter
\cite{quentinDM}.}
The variable $\CW_\CD := \langle{\cal R}\rangle_\CD - 6k_{\CD}/a_{\CD}^2$ describes the deviations of the average Ricci scalar curvature from a Friedmannian (eventually scale-dependent) constant curvature $k_\CD$ on the domain $\CD$, and the variable 
$\CQ_\CD$ is the backreaction functional, made up of averaged principal scalar invariants of the expansion tensor:
\begin{equation}
{\cal Q}_{\cal D}\;=\; 2\langle{\mathrm{II}}\rangle_{{\cal D}}-\frac{2}{3}\langle{\mathrm{I}}\rangle_{{\cal D}}^2 \ .
\end{equation}
Let us look at two examples. First, we look at the LTB (Lema\^\i tre-Tolman-Bondi) solution \cite{Lemaitre:1933gd,tolman1934effect,kras1,kras2}. Being an exact solution, the LTB models have been used as a reference to show that Zel'dovich-type approximations are more accurate than the standard linear and second-order perturbation theory (although the modeling of the velocity field is not as accurate as that for the density field) \cite{rzaltb}. 

Using the relation between the expansion tensor and the metric tensor in the coordinate form
$\Theta^i_{\;j}: =\frac{1}{2}g^{ik}\dot g_{kj}$,
the averaged scalar invariants of the expansion tensor on a simply-connected LTB-domain can be calculated \cite{rza2}:
\begin{equation}
\ltbaverage{\mathrm{I}(\Theta^i_{\;j})}=\frac{4\pi}{V_{LTB}}\int_0^{r_{\cal D}}\frac{\partial_r\left(\dot RR^2\right)}{\sqrt{1+2E}}\dd r \;;
\end{equation}
\begin{equation}
\ltbaverage{\mathrm{II}(\Theta^i_{\;j})}=\frac{4\pi}{V_{LTB}}\int_0^{r_{\cal D}}\frac{\partial_r\left(\dot R^2R\right)}{\sqrt{1+2E}}\dd r \;;
\end{equation}
\begin{equation}
\ltbaverage{\mathrm{III}(\Theta^i_{\;j})}=\frac{4\pi}{3V_{LTB}}\int_0^{r_{\cal D}}\frac{\partial_r\left(\dot R^3\right)}{\sqrt{1+2E}}\dd r \;,
\end{equation}
where $R$ and $E$ are functions of the coordinates $r,t$, and where the volume is given by
\begin{equation}
 V_{LTB}=\frac{4\pi}{3}\int_0^{r_{\cal{D}}}\frac{\partial_r\left(R^3\right)}{\sqrt{1+2E}}\dd r \;.
\end{equation}
These integrals can be straightforwardly  solved for $E(r)=E_0 = const.$ yielding for the averaged invariants:
\begin{equation}
\ltbaverage{\mathrm{II}(\Theta^i_{\;j})} = \displaystyle\frac{1}{3}\ltbaverage{\mathrm{I}(\Theta^i_{\;j})}^2\;\;;\;\;
\label{eq:spherical-I-III}
\ltbaverage{\mathrm{III}(\Theta^i_{\;j})} = \displaystyle\frac{1}{27}\ltbaverage{\mathrm{I}(\Theta^i_{\;j})}^3\,\;.
\end{equation}
This proves that, for a spatially Ricci-flat LTB model, 
the backreaction term vanishes.

Second, we calculated in \cite{rza2} the backreaction model provided by the \textit{Relativistic Zel'dovich approximation} (RZA), i.e. the first-order solution of the Lagrange-Einstein system:
\begin{subequations}
\label{Qfunctional}
\begin{eqnarray}
\label{resultQ2}
&^{\rm RZA}{\cal Q}_{\cal D}\;=
&\displaystyle\frac{\dot\xi^2\left(\gamma_1+\xi\gamma_2+\xi^2\gamma_3\right)}{\left(1+\xi\langle{\rm I}_{\rm \bf i}\rangle_{{\cal C_D}}+\xi^2\langle{\rm II}_{\rm \bf i}\rangle_{{\cal C_D}}+\xi^3\langle{\rm III}_{\rm \bf i}\rangle_{{\cal C_D}}\right)^2}\;\;,
\end{eqnarray}
where we have defined the set of initial data featuring the initial principal scalar invariants of the expansion tensor,
their initial average $\langle ... \rangle_{\cal C_D} := (1/V_{\CD}(t_i))\int_{\CD (t_i)} ... {\rm J}(X^i, t_i) {\mathrm d}X^i$, with ${\rm J}(X^i,t_i) = 1$, (the first term is the initial backreaction term):
\begin{eqnarray}
\label{Qinitial}
\gamma_1: = 2\langle{\rm II}_{\rm \bf i}\rangle_{{\cal C_D}}-\frac{2}{3}\langle{\rm I}_{\rm \bf i}\rangle_{{\cal C_D}}^2 = \CQ^{\rm initial}_{{\cal C}_D}\;;\nonumber\\
\gamma_2: = 6\langle{\rm III}_{\rm \bf i}\rangle_{{\cal C_D}}-\frac{2}{3}\langle{\rm II}_{\rm \bf i}\rangle_{{\cal C_D}}\langle{\rm I}_{\rm \bf i}\rangle_{{\cal C_D}} \;;\nonumber\\
\gamma_3: = 2\langle{\rm I}_{\rm \bf i}\rangle_{{\cal C_D}}\langle{\rm III}_{\rm \bf i}\rangle_{{\cal C_D}}-\frac{2}{3}\langle{\rm II}_{\rm \bf i}\rangle_{{\cal C_D}}^2 \;.
\end{eqnarray}
\end{subequations}
It is interesting to point out the leading, large-scale behaviour of this functional: taking as an example an Einstein-de Sitter background (with vanishing curvature and vanishing $\Lambda$), the leading term is $^{\rm RZA}{\cal Q}_{\cal D}\approx (1/a)\CQ^{\rm initial}_{{\cal C}_D}$. 
It corresponds to an approximation of the exact scaling solution of the averaged system, ${\cal Q}_{\cal D} = (1/a_\CD )\CQ^{\rm initial}_{{\cal C}_D}$, 
leading to the same scaling behaviour for the average spatial scalar curvature $\langle{\cal R}\rangle_\CD$, via the following coupling equation (following from the last equation of Eq.~\eqref{eq:effectivefriedmann}):
\begin{equation}
0=\frac{1}{a_\CD^{6}}\partial_t \left(\CQ_\CD  a_\CD^{6}\right)+\frac{1}{a_\CD^2}\partial_t\left(\langle {\cal R}\rangle_\CD a_\CD^2 \right)\ . \label{equ:integrab}
\end{equation}
This shows that the averaged curvature evolution can be very different from the evolution of a constant-curvature model, and this result can be profitably used to provide a fit to supernova data without dark energy \cite{asta1,asta2}. 

The backreaction functional \eqref{Qfunctional} was implemented numerically in the GPL-licensed code \cite{inhomog}.
The {\sc inhomog} library performs a numerical integration of the averaged Raychaudhuri equation for the domain-dependent volume scale factor using the RZA analytical formula for kinematical backreaction. 

As a further result, the RZA backreaction model also predicts vanishing backreaction in the case of the averaged Ricci-flat LTB model.
(For more details see \cite{buchert:focus}.)
A corresponding Newtonian derivation for the ZA model is provided in \cite{bks} and it arises from \eqref{resultQ2} and \eqref{Qinitial} through the dictionary \eqref{dictionary}. The result that only a restricted class of averaged LTB models is contained in the RZA backreaction functional  forms one of the motivations that led us to generalize the relativistic Lagrangian perturbation theory or the RZA model, to which we come later.

\begin{remark}
In Newtonian cosmology, the backreaction functional $\CQ_{{\mathbb T}^3}$ also vanishes for the deviations off a Friedmannian background cosmology by employing periodic boundary conditions, i.e. confining the deviations to a flat torus ${\mathbb T}^3$. This is mainly a result of the flatness of space and the fact that the principle scalar invariants of the expansion tensor are integrable, i.e. they can be written as divergences of vector fields as demonstrated in \cite{buchert94}. The backreaction functional can then be written as integrals over flux terms that have to vanish on a toroidal space $\mathbb{T}^3$.
The proof can be found in \cite{buchert:average}.
\end{remark}

Paper IV \cite{rza4} in the series gives a comprehensive investigation of the linearization of the tensorial perturbations, involving decompositions into trace and trace-free parts, where the transverse trace-free part is identified with Lagrangian gravitational waves. 
This paper also invokes Hodge theory to provide detailed insight into the integrable parts of the solution. 
It also provides a detailed dictionary of the variables if compared with standard perturbation theory, i.e. it shows which additional steps of linearization are involved to recover the Eulerian-linearized gravitational waves solutions, and it explains the relation of the Hodge decomposition to the standard Scalar-Vector-Tensor decomposition.
This paper also makes clear the role of the momentum constraints \eqref{form_momcoeff} for the interpretation of  the trace-free part of the Ricci tensor in \eqref{form_eomcoeff}: these latter are used to split the trace-free part of the Ricci tensor into gravito-electric and gravito-magnetic parts, where the former are tight to the sources (e.g. perturbations evolve according to the trace-solutions), and the latter to freely propagating waves in the linearized system. 

As an example, we here give the evolution equations for the gravito-electric and gravito-magnetic spatially projected parts of the Weyl tensor.
As ${E}_{ij}$ and ${H}_{ij}$ are symmetric and traceless tensors, at first order the constraints and evolution equations reduce to the following set: 
\begin{subequations}
\begin{eqnarray}   
   E_{ki\vert k}  &=&  - \tfrac{2}{3 a^3} \, W_{\:\vert i}\:;\label{divElin}   \\
   H_{ki \vert k}   &=& 0 \:; \label{divHlin} \\
   \dot{E}_{ij} +3H {E}_{ij}  - \tfrac{1}{a} \epsilon_{(i}{}^{\,\,kl}H_{j)l \vert k}   &=& - 4 \pi G {\varrho}_H\, \sigma_{ij} \:; \quad\label{rotHlin} \\
   \dot{H}_{ij} +3 H {H}_{ij}  + \tfrac{1}{a} \epsilon_{(i}{}^{\,\,kl} E_{j)l \vert k}   &=& 0\label{rotElin} \:,\\ \nonumber
\end{eqnarray}
\end{subequations}
where ${\varrho}_H =({\varrho}_H)_{\rm\bf i}\,/\,a^3 $ denotes the homogeneous part of the density field, $W = -4\pi G \varrho_H \delta$ is related to the density perturbation $\delta$, and $\sigma_{ij}$ are the components of the shear tensor.
We obtained the following second-order propagation equations: 
\begin{subequations}
\begin{equation}
\begin{split}
\label{Efinal}
    \Box_0 E_{ij}  - 7 H \dot{E}_{ij} - 4 (4\pi G {\varrho}_H+ \Lambda) E_{ij}
    = - \frac{1}{a^5} \mathcal{D}_{ij} W  + 4 \pi G H {\varrho}_H \,\sigma_{ij} \ ;
    \end{split}
\end{equation}
\begin{equation}
\begin{split}
\label{Hfinal}
 \Box_0 {H}_{ij} - 7 H \dot{H}_{ij} - 4 (5\pi G {\varrho}_H+ \Lambda) H_{ij} =  - \frac{1}{a}4 \pi G {\varrho}_H \epsilon_{(i}{}^{\: kl} \sigma_{j)l\vert k} \ ,
    \end{split}
\end{equation}
\end{subequations}
where $\Box_0 {X}_{ij} : = -\ddot{X}_{ij} + \tfrac{1}{a^2} \Delta_0 {X}_{ij}$ denotes the Lagrangian d'Alembertian applied to the tensor field ${X}_{ij}$.\footnote{The coefficients in front of the third terms in Equations \eqref{Efinal} and \eqref{Hfinal} differ, since the first equation
features a term $\propto {\dot\sigma}_{ij}$ that can be replaced through ${\dot\sigma}_{ij} = - E_{ij}$, \textit{cf.} Equation (111) in \cite{rza1}, which changes the coefficient $5\pi G {\varrho}_H+ \Lambda$ to $4\pi G {\varrho}_H+ \Lambda$.}

This work also illustrates the inherently nonlinear character of a Lagrangian linearization, in contrast to a linearized propagation of gravitational waves written in global background coordinates, as was mentioned in the introduction: gravitational waves are perturbations of space and they propagate within the perturbed space in a Lagrangian-linear approach. 

Paper V \cite{rza5} in the series takes pressure into account. Contrary to the Newtonian treatment of a perfect fluid source \cite{adlerbuchert}, the general-relativistic approach needs to account for a foliation change, invoking a lapse function in the simplest case \cite{buchert:fluid}. This aspect is also interesting compared with the approach of standard perturbation theory in cosmology: pressure-supported fluids like a radiation fluid are treated within the same foliation of spacetime, while an intrinsic description of a dust and a radiation fluid, respectively, physically live on different foliations. Since substantial information is drawn from Cosmic Microwave Background radiation properties -- such as the ratio of peaks in the spectrum of baryon acoustic oscillations -- and compared with late-time structure formation in a dust matter model, such issues might alter those interpretations and it would be worthwhile to dive deeper into this difference.

\section{Non-perturbative generalization of Relativistic Lagrangian perturbation theory}

Paper VI \cite{rza6} in the series takes its motivation from the remarks of the spatially averaged LTB model above, but also from an early insight of Masumi Kasai \cite{kasai95} that the relativistic form of Zel'dovich's approximation contains some of the exact Szekeres solutions in a subcase. We face the property that Lagrangian perturbation solutions at first order (RZA) provide an exact description in the maximally anisotropic case, while, on average, the opposite behavior is also contained in a subclass, namely the averaged LTB model. The paper VI \cite{rza6} headed at making the link to the exact class of Szekeres solutions precise with the main results that (i) Szekeres Class II solutions are contained in a subclass of RZA and they furnish exact deviations from a global FLRW background cosmology for all times, and that (ii) the spatial average of these exact deviations has to vanish. The latter result reveals a crucial restriction of RZA: vanishing averages of deviations from a global FLRW background imply that these models cannot describe global backreaction, i.e. the inhomogeneous deviations evolve according to the chosen background cosmology,
but the background cosmology is not affected by the inhomogeneities. This construction implies a conservation law for the average scalar curvature in conformity with the background curvature evolution, which is not the case in general \cite{buchertcarfora2}: changes are due to the coupling of structure formation to the average scalar curvature according to the general conservation law Eq.~\eqref{equ:integrab}. This remark is equally relevant for general-relativistic numerical simulations that are globally constructed in the same way: the evolution of the global scalar curvature can be used to test as to whether the simulation allows for global deviations from the background curvature model, or whether its architecture is, by construction, limited to conserve the imposed background curvature. 

In view of these remarks it is natural to ask whether we could generalize the relativistic first-order solutions and their averages in such a way that, locally, also Szekeres Class I solutions are contained in a subclass and hence also LTB models with curvature. 
Before we come to the proposed generalization, we shall look into the family of Szekeres solutions.

\subsection{Szekeres solutions in the Goode and Wainwright parametrization}

The family of solutions found by Szekeres \cite{Sz75} and later generalized by Szafron \cite{szafron1977inhomogeneous} constitutes the most general exact solution to Einstein's equations that can be regarded as an inhomogeneous cosmological model. 
The general line-element can be written in the form\footnote{Henceforth, we use units where $c=G=1$.} \cite{kras1,kras2}: 
\begin{equation}\label{Eq:GeneralSzeMetric}
\dd s^2=-\dd t^2+e^{2 \alpha} \dd \z^{2} + e^{2 \beta} \left(\dd \x^{2}+\dd \y^{2}\right),
\end{equation}
where the metric functions $\alpha$ and $\beta$, in general, depend on all the comoving coordinates. 
The field source of the Szekeres solution is a perfect fluid with inhomogeneous density but an only time-dependent pressure $p=p(t)$, which for physical models reduces to an irrotational but inhomogeneous dust fluid with the cosmological constant as the dark energy model \cite{SzaSpatialConfFlat1978} -- see \cite{CommentSzePress}
for a discussion on the limitations imposed by this property. 

In these coordinates, the solution naturally splits into two classes: 
class I when $\beta_{,\z}\neq0$  and class II when $\beta_{,\z}=0$. In general, neither class has symmetries 
\cite{SzNoSymm1}, and other well-known exact solutions are contained as particular cases, e.g., the LTB model follows from the 
class I, and some subclasses of the Bianchi models emerge from the homogeneous limit of class II. In general, both classes have a well-defined FLRW limit. Szekeres models have been used to examine several astrophysical and cosmological problems, being the first class the one that has caught the most attention \cite{hellaby1996null,HellabyKrasi2002,Bolejko2006Struformation,bolejko2007evolution,WaltersHellaby2012,Buckley2013,Vrba:2014,Koksbang2015,Koksbang2015II,Koksbang2017,Coley:2019ylo,Hellaby:2007hq,IshakNwankwo2008,Bolejko:2009GERG,BolCelerier2010,BolSuss2011,SussBol2012,sussman2015multiple,sussman2016coarse,gaspar2018black}. 

The original solution has been reparametrized multiple times. Among these para\-metrizations, the Goode and Wainwright's \cite{GW1} is the one that provides the most direct connection with RZA. 
In this representation, the line-element~\eqref{Eq:GeneralSzeMetric} is rewritten as \cite{kras1,kras2}
\begin{equation}\label{Eq:SzeMetricGW}
\dd s^2=-\dd t^2 + \s^2  \left(\GG ^2 \W^2 \dd \z^{2} + e^{2 \nu}  \left( \dd \x^{2} + \dd \y^{2}\right) \right) \ ,
\end{equation}
where the scale factor $\s$ obeys a Friedmann-like equation:
\begin{equation}\label{Eq:FriedmannLikeEqn}
\left(\frac{\dot{\s}}{\s}\right)^2=-\frac{k_0}{\s^2} +  \frac{2\mu}{\s^3} +\frac{\Lambda}{3}\ ,
\end{equation}
with $k_0=0,\pm1$ and $\mu>0$ being an arbitrary function. Also,
\begin{equation}\label{Eq:GDef}
\GG=\A(\rv)-\F(t,\z) \quad \hbox{with} \quad \F:=\beta_{+} f_{+}+\beta_{-} f_{-} \ .
\end{equation}
One of the most enlightening features of this parametrization is that $f_{\pm}$ (and then $\F$) satisfy the same equation that leads to the growing and decaying modes in cosmological perturbation theory:
\begin{equation}\label{Eq:Ftt}
\ddot{\F}+ 2\frac{\dot{\s}}{\s}\dot{\F}-\frac{3\mu}{\s^3}\F=0 \ .
\end{equation}
The functions $\mathcal{A}(\rv)$, $e^{2\nu(\rv)}$, $\mathcal{W}(\z)$, $\beta_{+}(\z)$ and $\beta_{-}(\z)$ differ for each class, see their specific form in \cite{GW1,kras2} or \cite{rza6}, the latter for a presentation using the notation used in this paper. Here we restrict ourselves to highlight the most important characteristics of each class:
\begin{itemize}
    \item  {\bf Class I:} 
    
    $\s=\s(t,\z)$, $\mu=\mu(\z)$, 
    $f_{\pm}=f_{\pm}(t,\z)$, $\beta_+=-k_0 f \frac{\mu_{,\z}}{3\mu}$ and  $ \beta_-=f  \mathcal{T}_{,\z} $ where $\mathcal{T}(\z)$ is the time of the initial singularity and $f(\z)$ is an arbitrary function.
    
    \item {\bf Class II:}
  
    $\s=\s(t)$, $f_{\pm}=f_{\pm}(t)$, $\mu=\hbox{const.}$,  $\W=1$. 
\end{itemize}

\noindent
Regardless of the class, the energy density can be integrated to the following compact expression:
\begin{equation}\label{Eq:Rho2ClassesGenEq}
 4\pi\varrho(t,\rv)=\frac{3\mu}{\s^3}\left(1+\frac{\F}{\GG}\right) \ .
\end{equation}

\subsection{Szekeres models and Lagrangian perturbation theory}

The starting point to relate the Lagrangian formalism and Szekeres exact solutions is the comparison of their line-elements. 
We first identify the initial metric ($G_{ij}$) and the exact deviation 
(${h}_{ij}$):
\begin{equation}
g_{ij}\equiv a^2 (t) \gamma_{ij} \equiv a^2 (t) \left(G_{ij} 
+ {h}_{ij} \right) \ ,
\label{RZAansatzSzeII}
\end{equation}
and then find the Szekeres deformation field by solving the equations from the equality of the Szekeres metric and the bilinear functional of RZA: 
\begin{equation}\label{GammaijSze} 
{h}_{ij}\equiv G_{ab} \left(\delta^{a}_{\ i}  P^{b}_{\ j} + \delta^{b}_{~j} P^{a}_{\ i}  + P^{a}_{~i} P^{b}_{\ j} \right) \ .
\end{equation}
Finally, we compare the evolution equation(s) for the thus defined Szekeres deformation field with the model equations of RZA. 

At this point, it is convenient to split the analysis into the two different classes and treat them separately.

\subsubsection{Class II models as the exact limit of RZA}

For this class, the parameter $\mu$ is a constant, and the conformal scale factor is a function of the time only. If we set $\mu$ equal to the energy density of the homogeneous
universe model at the initial time ($3\mu=4\pi \varrho_b(t_i)$), Eq. \eqref{Eq:FriedmannLikeEqn} is precisely the Friedmann equation.
Below, we will relabel $\s=a(t)$ and assume that $a(t)$ is initially normalized: $a(t_i)=1$.
The initial metric is via Gram's matrix determined by the direct evaluation of \eqref{Eq:SzeMetricGW} at $t=t_i$:
\begin{equation}\label{Eq:GabClassIISol}
G_{ab}=\hbox{Diag}\left[\widetilde{\A}^2, e^{2\nu},e^{2\nu} \right] \quad \hbox{with} \quad \widetilde{\A}:= \A-\F_i \quad \left(\hbox{and} \quad \F_i:=\F(t_i)\right)\ .
\end{equation}
Comparing the line-elements~\eqref{Eq:SzeMetricGW} and~\eqref{GammaijSze}, 
we obtain that the only non-vanishing component of $h_{ij}$ is ${h}_{33}$ (or ${h}_{\z\z}$), which is given by
\begin{equation}
{h}_{33} =-2 \A \widetilde{\F} +\widetilde{\F}^2
= {G}_{ab} \left(\delta^{a}_{~3}  P^{b}_{~3} + \delta^{b}_{~3} P^{a}_{~3}  + P^{a}_{~3} P^{b}_{~3}\right)  = 2 \widetilde{\A}^2 P^{3}_{\ 3} +  \widetilde{\A}^2 \left( P^{3}_{\ 3} \right)^2 \ , 
\end{equation}
where we have defined $\widetilde{\F}=\F-\F_i$. Thus, $P^{3}_{\ 3}$ is the only non-trivial component of the deformation field:
\begin{equation}\label{Eq:P33eqn1}
P^{3}_{\ 3} =P= -\widetilde{\F}/\widetilde{\A} \ ,
\end{equation}
furnishing a locally one-dimensional solution.

It can also be verified by direct substitution that $P$ obeys the evolution equation for the trace part of the 
deformation field -- the only non-trivial one for Szekeres and equivalent to Eq.~\eqref{Eq:Ftt} for Szekeres models.
Also, all Szekeres fields are reproduced by the functional expressions of RZA.
This analysis can be summarized in the following remark.

\begin{remark}
The class II of Szekeres models is contained in RZA within its locally one-dimensional case, and the dynamics is exclusively governed by the first principal scalar invariant of the deformation field.
Applying the dictionary \eqref{dictionary} we find the exact 3D solutions without symmetry in Newtonian cosmology \cite{buchert89}.
\end{remark}

Hence, Szekeres class II emerges from a proper set of the initial data, for which the constraints are satisfied on the initial hypersurface. This result illustrates the nonlinear nature of RZA. 
We here note, in particular, that the metric is a bilinear functional of the linearized perturbation, hence quadratic, and the density is given by the exact integral of the continuity equation with the second and third principal scalar invariants vanishing. As for the averages, since the background model is global for this class, the deviation fields lead to a vanishing backreaction functional for periodic boundary conditions \cite{rza6}.

Although these solutions (Szekeres II and its Newtonian analog) are valuable for understanding the theory, their applications as cosmological models are limited by the difficulty of finding initial data without singularities. In Szekeres II models with a flat associated background, the presence of a growing mode necessarily leads to singularities on the initial hypersurface. Still, these models have found cosmological applications considering only the regular part of the 
manifold~\cite{Meures-Bruni_mnras,Meures-Bruni_prd}.  

\subsubsection{Class I models and  Lagrangian perturbation theory}

The relationship with the Lagrangian perturbation theory is more involved for class I than for class II models. 
Since class I lacks a global background, all relevant dynamical functions ($\s$ and $f_\pm$) exhibit a nontrivial dependence on the spatial coordinates, making it impossible to separate the space- and time-dependencies. 

Still, we consider the co-frame set as the fundamental dynamical field and retain the mathematical structure of the Lagrangian perturbation theory. In this way, the field $\hat{P}^a_{\ i}$ is reinterpreted as a local deformation field on a Friedmann-like local `reference model' described by Eq.~\eqref{Eq:FriedmannLikeEqn}.\footnote{
To avoid confusion, we will use $\hat{P}^a_{\ i}$ to refer to the  deformation field with respect a generic reference model and reserve $P^a_{\ i}$ for the deformation with respect to the FLRW background. } 
Following similar arguments to those used for class II, it can be shown that such a deformation field has only one non-trivial component, which has the same functional expression as its equivalent for class II, Eq.~\eqref{Eq:P33eqn1}. 

To dig more deeply into the connection of RZA and the exact solutions, we exploit the \textit{silent property} of Szekeres, under which each world line evolves ``independently''  of the others~\cite{BruniPantano1994,vanElst:1995,Bolejko:2017SilentUniv1}. 
Let us consider an arbitrary world line labeled by the comoving coordinates $\rv_\ast$. 
In the evolution of this fluid line, the scale factor
\begin{equation}
   \hat{a}(t):=\s(t,\z)|_{\z=\rv_\ast}  
\end{equation}
obeys the Friedmann equations for a model with initial density $4\pi \varrho_b(t_i) =3 \mu(\z)|_{\z=\rv_\ast}$: the `reference model'. 
Then, the local deformation field introduced in the above paragraph satisfies the RZA evolution equation for its trace part (the only non-trivial one).
As for class II, this evolution equation reduces to~\eqref{Eq:Ftt}, and  $f_\pm(t,\z)|_{\z=\rv_\ast}$ are interpreted as the growing and decaying modes on the `reference model'.

\begin{remark}
The dynamics of the Szekeres models of class I can be interpreted as a constrained superposition of world lines satisfying the RZA model equations sourced by different ``backgrounds'', i.e., a local Friedmann-like reference model.
All the local quantities characterizing the world lines can also be obtained from their RZA functional expressions.
Averaging over these ``local evolutions'' will provide a new global average model that interacts with the inhomogeneities, evolves differently compared with a Friedmannian model and, hence, includes cosmological backreaction. 
\end{remark}

This analysis paves the way for a strategy to generalize RZA so that it can reproduce the entire family of Szekeres solutions (and thus the LTB models) for a proper set of the initial data.


\subsubsection{Rescaled class I solution}\label{SubSec:rescsol}

To examine the relation with RZA it is convenient to normalize the initial conformal scale factor. 
While for class II this can be easily done by a re-scaling of the comoving coordinates, 
for class I the initial scale factor is a function of $\z$: $\s(t_i)=\s_i(\z)$. 
Then, we define $\As:=\s/\s_i$, which clearly satisfies $\As(t_i)=1$. Proceeding along these lines, Eq.~\eqref{Eq:FriedmannLikeEqn}  keeps the form of a Friedmann equation but in terms of space-dependent curvature and density parameters, $\hat{k}(\z):=k_0/S_i^2$ and $4\pi\hat{\varrho}(\z):=3 \mu/S_i^3$:
\begin{equation}\label{Eq:FriedLikeAs}
    \left(\frac{\dot{\As}}{\As}\right)^2=-\frac{\hat{k}}{\As^2} + \frac{8 \pi}{3} \frac{\hat{\varrho}}{\As^3} +\frac{\Lambda}{3} \ ,
\end{equation}
and $\F$ remains as the superposition of the growing and decaying modes on the `reference model' defined by~\eqref{Eq:FriedLikeAs}:
\begin{equation}\label{Eq:FttAs}
\ddot{\F}+ 2\frac{\dot{\As}}{\As}\dot{\F}-\frac{4\pi \hat{\varrho}}{\As^3}\F=0 \ .
\end{equation}
Other minor changes to the arbitrary functions are also needed to leave the solution unaltered.


\subsection[Generalized RZA:  an exact-solution controlled model of structure formation]{Generalized RZA:  an exact-solution controlled model of structure formation\footnote{In this section, we highlight some of the salient results of a detailed investigation \cite{GRZA1}.}}

Although RZA cannot reproduce the class I of Szekeres models, both solutions are governed by the same set of evolution equations. Their difference purely relies on the constraints. 
This is a characteristic feature of the silent class of exact solutions to Einstein's equations, where some local rotational symmetric and Bianchi solutions and the LTB models have a common set of evolution equations. It is precisely the constraints that make them different \cite{Bolejko:2017SilentUniv1}.

Motivated by these results, we propose a generalization of RZA with as minimal modifications as possible to contain Szekeres class I as a particular case, while still retaining the mathematical structure of the Lagrangian perturbation theory. 
Proceeding along these lines, the co-frames are kept as the fundamental field of the approach. 
Next, we introduce a Friedmann-like `reference model' as a generalization of the homogeneous background and the deformation field on it,  
\begin{equation}
\bm{\eta}^a=\eta^a_{ \ i} \mathbf{d}X^i =\As \left(\delta^{a}_{\ i}+\hat{P}^{a}_{\ i}\right) \mathbf{d}X^i \ ,
\quad
 \As=\As\left(t,\rv \right) \ , \quad  \hat{P}^{a}_{\ i}=\hat{P}^{a}_{\ i}\left(t,\rv\right) \ .
\end{equation}
This `reference model' satisfies Eq.~\eqref{Eq:FriedLikeAs} with a general curvature ($\hat{k}=\hat{k}\left(\rv\right)$) and density parameter ($\hat{\varrho}(\rv)$). 

The subsequent approach consists of inserting this co-frame set into the Lagrange-Einstein system \eqref{LES} and linearizing the evolution equations in the deformation field. As in RZA, this formal solution for $\hat{P}^{a}_{\ i}$ is injected into the nonlinear functional expressions for the physical fields, e.g., density, 3D Ricci scalar, expansion scalar and the (bilinear) metric. We call this approach \textit{Generalized RZA} (GRZA).

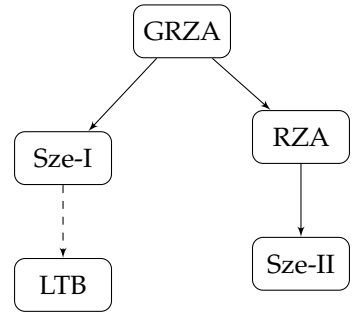
\begin{figure}
\begin{center} 
\tikzstyle{block} = [rectangle, draw,   
    text width=3em, text centered, rounded corners, node distance=4.8em, minimum height=2em]  
\tikzstyle{line} = [draw, -latex']  
\tikzstyle{arrow} = [thick,->,>=stealth]
\begin{tikzpicture}[node distance = 10em, auto] 

\node [block](GRZA){GRZA}; 
\node [block, below of = GRZA, xshift=-4.5em](SzeI){Sze-I};  
\node [block, below of = GRZA, xshift=4.5em,yshift=.8em](RZA){RZA};  
\node [block, below of = RZA](SzeII){Sze-II};  
\node [block, below of = SzeI](LTB){LTB};
      
 \path [line](GRZA) -- (SzeI); 
 \path [line](GRZA) -- (RZA);    
  \path [line](RZA) -- (SzeII); 
   \path [line,dashed](SzeI) -- (LTB); 
\end{tikzpicture}  
\end{center}  
\caption{Special subcases of GRZA.
GRZA contains class I of the Szekeres (Sze-I) solutions and all its subcases, such as the LTB models. Also, RZA and the class II of Szekeres (Sze-II) emerge in the limit when $\hat{k}(\rv)\rightarrow k_0=\hbox{const.}$ and $\hat{\varrho}(\rv)\rightarrow \varrho_b(t_i)=\hbox{const.}$}
\label{DiagramGRZA}
\end{figure}

It is instructive to rephrase the previous analysis in different terms. 
GRZA results from adding two extra degrees of freedom to RZA; namely,
the constant curvature and density of the FLRW model are generalized by space-dependent functions:
\begin{equation}
\hbox{RZA} \rightarrow \hbox{GRZA}:   
\qquad  k_0 \rightarrow  \hat{k}(\rv) 
\quad  \land \quad \varrho_b(t_i) \rightarrow  \hat{\varrho} (\rv) \ .
\end{equation}
This architecture allows reproducing the class I of the Szekeres solutions and its 
well-known particular cases, e.g. the LTB models.  
Also, RZA  and the class II of Szekeres emerge in the limit when the curvature and density parameters of the reference model become constant, see Fig. \ref{DiagramGRZA}.

\subsubsection{An example of GRZA numerical simulation}

GRZA has enormous potential to model the large-scale structure formation in general relativity without suffering from the constraints of a global background architecture. As a proof of principle, we show a simple numerical example obtained from a set of Szekeres-like initial data at the last scattering time. 
This family of models is characterized by a locally one-dimensional deformation field, where the Szekeres exact expression is modified to include a fluctuation proportional to the growing and decaying modes: 
\begin{equation}
\hat{P}(t,\rv)=\frac{\hat{\F}-\hat{\F}_{i}}{\A-\hat{\F}_{i}} \ ,
\quad
\hat{\F}=\F+\delta \F \ ,\quad \delta \F=\delta \beta_+ f_+ + \delta \beta_- f_-  \ .
\end{equation}
In the equation above, $\F=\beta_+ f_+ + \beta_- f_-$ is the Szekeres function introduced in \eqref{Eq:GDef}. 

\begin{figure*}[ht]
     \centering
     \begin{subfigure}[b]{0.32\textwidth}
         \centering
         \includegraphics[width=\textwidth]{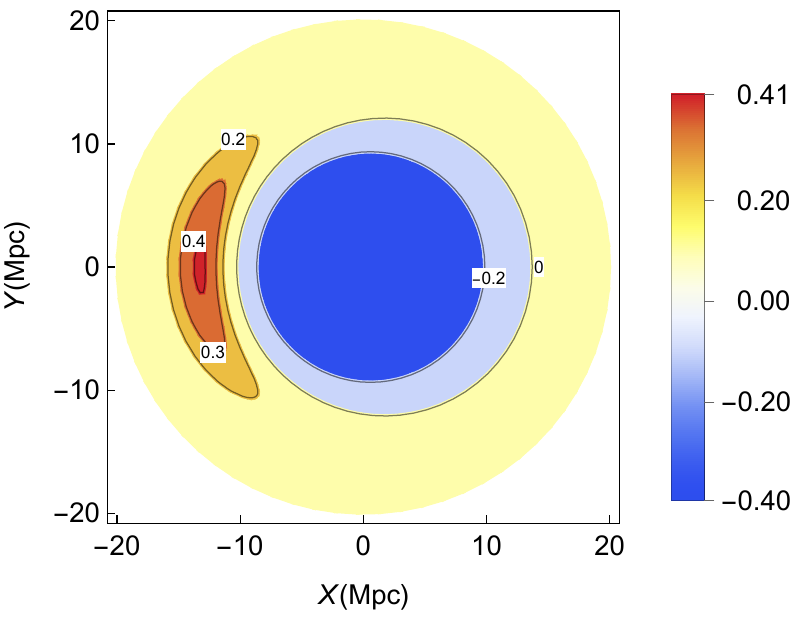}
         \caption{$\alpha=0$}
         \label{fig:alp0}
    \end{subfigure}
    \begin{subfigure}[b]{0.32\textwidth}
         \centering
         \includegraphics[width=\textwidth]{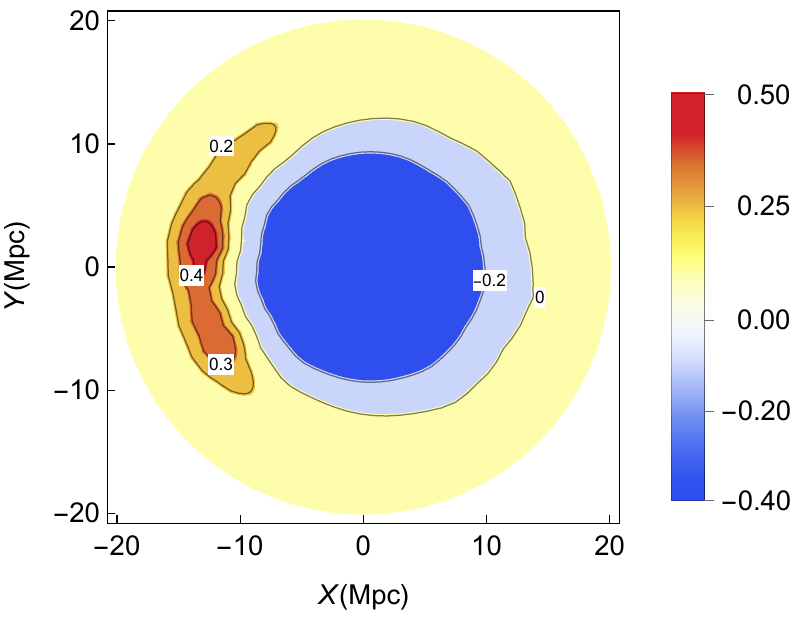}
         \caption{$\alpha=8.9\times 10^{-4}$}
         \label{fig:alp1}
     \end{subfigure} 
    \begin{subfigure}[b]{0.32\textwidth}
         \centering
         \includegraphics[width=\textwidth]{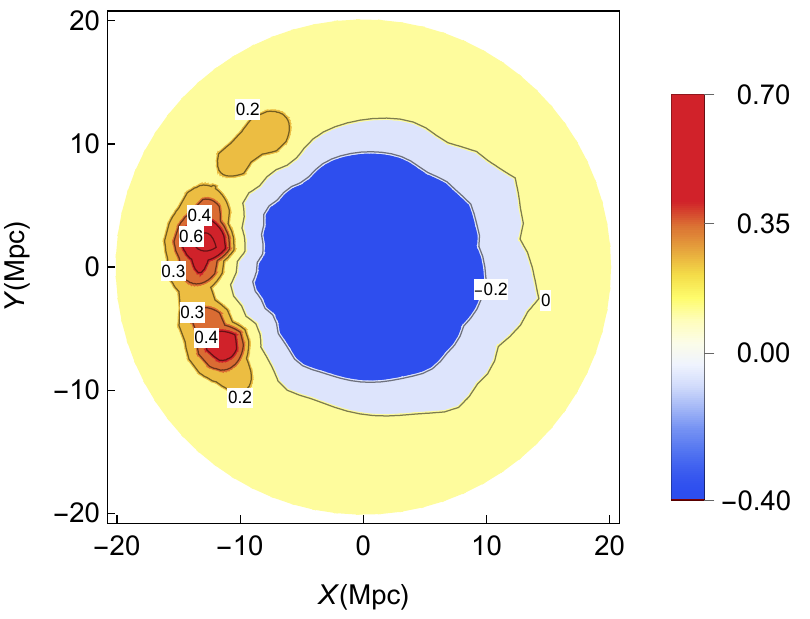}
         \caption{$\alpha=2.2\times 10^{-3}$}
         \label{fig:alp2}
     \end{subfigure}
     \vspace{10pt}
     \caption{Density contrast in the equatorial plane at the present cosmic time. 
     The case $\alpha=0$ corresponds to an exact Szekeres model; then, the Szekeres pancake-shaped overdensity splits into three structures.
     The coordinates are defined as $X= \As(t_0,\z) \z\cos\phi$ and $Y=\As(t_0,\z) \z \sin\phi$.}
     \label{fig:snapshots}
\end{figure*}

\begin{figure}[ht]
\centering
\includegraphics[width=0.80\textwidth]{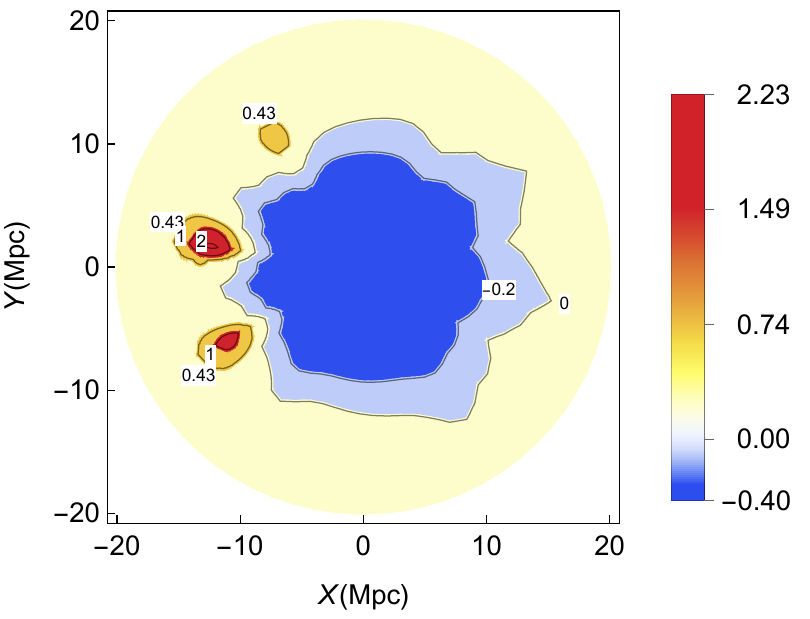}
\caption{Density contrast in the equatorial plane at the present cosmic time for $\alpha=5.2\times 10^{-3}$.
The coordinates are defined as in Fig. \ref{fig:snapshots}.}
\label{fig:alp3}
\end{figure}

As thus built, these models retain the characteristic foliation of the hypersurfaces by non-concentric two-spheres of the quasi-spherical models, but generalize the Szekeres solution in the way these 2-spheres are arranged. Figures \ref{fig:snapshots} and \ref{fig:alp3} show the present-day projection of the density contrast in the equatorial plane for different solutions with different deviations. 
The function $\delta \beta_+$ is assumed to have the form $\beta_+ \propto \alpha \sin^3(\gamma \pi \y) \sin^4(\pi \z)$; 
hence, the case $\alpha=0$ corresponds to the exact solution. 
While the models with $\alpha\neq 0$ are not exact, the violation of the constraints remains small throughout the entire evolution, indicating that they are valid approximate solutions of general relativity. (For these simulations the largest value of the violation of the constraints at the present cosmic time is of the order of $10^{-3}$ for the case depicted in Fig. \ref{fig:alp3}.)

Figure \ref{fig:snapshots} illustrates the breaking of the Szekeres ``dipolar symmetry''. As can be seen in the sequence Fig. \ref{fig:alp0}, \ref{fig:alp1}, and \ref{fig:alp2}, the characteristic Szekeres pancake-shaped overdensity splits into three structures (at the present cosmic time) as the value of $\alpha$ increases. Then, Fig. \ref{fig:alp3} shows the density contrast for a larger value of $\alpha$, leading to a network of structures undergoing non-linear collapse. 
To achieve a better visualization, in this graph we have used a different color scale than the one used in Fig. \ref{fig:snapshots}. ---The scale is shown in the right bar of each graph.

\bigskip

\vspace{10pt}

\newpage

\section{Summary, Discussion and Outlook}

After a historical review on the development of the Lagrangian perturbation theory in Newtonian cosmology, we recalled the Lagrange-Newton system \eqref{LNS1} governing the deformation gradient ${\bf d}f^a$, which is the starting point of a Lagrangian perturbation expansion on a global background cosmology. Since the deformation gradient is the only dynamical variable, we are entitled to calculate any field variable through its exact definition for a given order of the perturbative expansion.

Through the introduction of a simple dictionary, which replaces the exact one-form fields ${\bf d}f^a$ by general co-frame fields $\boldsymbol{\eta}^a$, we obtain a general-relativistic form of the Lagrangian deformation and its perturbative expansions. 
We recalled that the same logic applies to the Einstein equations, exemplified for an irrotational dust continuum in a flow-orthogonal foliation of space-time, namely that the Einstein equations can be represented through the Lagrange-Einstein system \eqref{LES} for the co-frame fields as the only dynamical variable. Consequently, we are entitled to calculate any variable in terms of exact functionals of the co-frame fields, e.g. the metric as a bilinear quadratic form of the co-frames, the expansion and Ricci tensors, etc., valid for any order of the perturbative deformation that we consider.
We remark that the Lagrange-Einstein system can be generalized for general energy-momentum tensors and general foliations of space-time, including tilted foliations (for a comprehensive investigation and the definition of the Lagrangian picture in the general situation, see \cite{buchert:generalfluid} and \cite{foliations}).  

We reviewed the main contents of a series of papers on the general-relativistic form of the Lagrangian perturbation theory on a global background cosmology. In particular, we also considered average properties of the Lagrangian first-order solutions that gave rise to a compact backreaction functional capturing the deviations from the assumed background cosmology. We listed a few known theorems: backreaction vanishes for Newtonian cosmological models that employ periodic boundary conditions. The same applies for spherically symmetric models on a flat space section. These results carry over to corresponding theorems in general relativity, i.e. for simulations that impose the torus architecture on flat space sections, and for flat LTB models. 
In a next step we looked at the class of Szekeres solutions, and we have shown that Szekeres Class II is contained in a subclass of the first-order relativistic Lagrangian perturbation solutions (RZA model). Integral constraints can be imposed to comply with the torus architecture of relativistic simulations that are conceived as deviations from a global background cosmology. 

We then motivated a non-perturbative generalization of the Lagrangian perturbation theory with the aim to also include Szekeres Class I and general LTB models.
This could be achieved by using the Goode-Wainwright parametrization of the Szekeres solutions, since this allowed to represent the solutions in terms of a Friedmann-like equation for the ``background'' and deviations thereof that also comply with the form of the perturbation equations known in Lagrangian perturbation theory. We presented the results that (i) Szekeres Class II solutions with a homogeneous scale factor correspond to the general-relativistic form of the Lagrangian first-order perturbation solutions, while (ii) Szekeres Class I solutions with an inhomogeneous scale factor still allow us to employ the Lagrangian formalism, but only locally. 

We gave a numerical example of a Szekeres Class I model on the scale of a collapsing structure in comparison with initial data that were slightly deformed within the class of the corresponding Lagrangian perturbation solution. These deformations we controlled via the condition of smallness of violation of the energy constraint. 

To proceed further towards a generic situation for the formation of large-scale structure in the Universe, we have to implement a local background and a corresponding deviation dynamics that is sourced by this local background evolution for a large-scale simulation. As this construction is local, we have to average the resulting dynamics to obtain the physical average cosmology. 
By construction, the resulting generalization of the Lagrangian model (GRZA) contains Szekeres Class I solutions in a subclass, but it also provides an exact-solution controlled approximation that features an evolving average distribution. 
The result is a model that is able to quantify global cosmological backreaction. It thus realizes the expectation that fluctuations ``talk'` to the ``global background'', conceived as the average model. 

Further investigations are needed to numerically realize this new approximation. We point out that an alternative implementation is possible: instead of locally splitting the dynamics into a ``local background'' and deviations thereof,
we may directly use the averaged variables such as the backreaction functional. This approach is quasi-local in the sense that average properties are implemented on a domain of the size of the resolution scale that we choose. The limit where the averaging domain $\CD$ tends to zero agrees with the local dynamics, so that we may apply averages to sufficiently small domains, then summing up the quasi-local results by using the exact volume-partitioning formula for the backreaction functional \cite{buchertcarfora2}, and the exact volume-partitioning formulas of a multiscale dynamics \cite{multiscale}. Control on this procedure is provided by the spatially averaged constraints of general relativity. 

A further alternative that is independent of the here proposed strategy is iterative rather than perturbative in nature: we can set up the deviation dynamics on a large-scale or ``global background model'' that is identified with the exact average dynamics. This approach has been outlined in \cite{generalbackground}.

\vspace{160pt} 



\authorcontributions{TB: conceptualization, writing, methodology, supervision, funding acquisition; IDG: writing, methodology, software, visualization; JJO: writing, funding acquisition, software supervision.
All authors have read and agreed to the published version of the manuscript.}

\funding{This work is part of a project that has received funding from the European Research Council (ERC) under the European Union's Horizon 2020 research and innovation programme (grant agreement ERC advanced grant 
740021--ARTHUS, PI: TB). IDG and JJO acknowledge support from the National Science Centre (NCN, Poland) under the Sonata-15 research grant UMO-2019/35/D/ST9/00342. IDG acknowledges hospitality and support by ERC ARTHUS during a working visit in Lyon.}

\dataavailability{The {\sc inhomog} code referred to in the context of realizing the backreaction functional in RZA is openly available \cite{inhomog}. A pedagogical jupyter notebook to realize the example for GRZA in comparison with Szekeres Class I solutions is available at 
\url{https://github.com/idgaspar/Beyond_relativistic_Lagrangian_perturbation_theory}.}


\acknowledgments{We would like to thank the guest editors Giovanni Marozzi and Luigi Tedesco for inviting us to contribute to this special issue. We would like to emphasize the fruitful inspirations that we got from Maurizio Gasperini's work on covariant averaging formalisms that he investigated with Gabriele Veneziano and collaborators (not being the main subject of this review).\\
TB wishes to thank Oliver Hahn and Cornelius Rampf for valuable discussions during a visit to Vienna Observatory, Austria. We are thankful to a referee for insightful comments and suggestions.}

\conflictsofinterest{
The authors declare no conflict of interest.}

\begin{adjustwidth}{-\extralength}{0cm}
\newpage
\reftitle{References}



\newcommand\eprintarXiv[1]{\href{http://arXiv.org/abs/#1}{arXiv:#1}}
\end{adjustwidth}
\end{document}